\documentclass[onecolumn]{aa}       
\usepackage{graphicx}
\usepackage{txfonts}
\usepackage{bm}

\begin{document}
\def\bea{\begin{eqnarray}}
\def\eea{\end{eqnarray}}
\def\be{\begin{equation}}
\def\ee{\end{equation}}
\def\rra{\right\rangle}
\def\lla{\left\langle}
\def\le#1{\label{eq:#1}}
\def\re#1{\ref{eq:#1}}
\def\eps{\epsilon}
\def\sgm{\Sigma^-}
\def\la{\Lambda}
\def\kv{\bm{k}}
\def\vd{{\cal V}} 
\def\pp{{\cal P}} 
\def\kk{{\cal K}} 

\title{Structure of protoneutron stars within a static approach}

\author{O. E. Nicotra}

\institute{Dipartimento di Fisica, Universit\`{a} di Catania and Istituto Nazionale 
di Fisica Nucleare, Sezione di Catania, Via S. Sofia 64, I-95123 Catania, Italy}

\date{Received / Accepted}


\abstract{
To investigate the stability of 
   protoneutron stars in their early evolution, the minimum gravitational mass
   plays a fundamental role. This quantity depends upon the temperature profile 
   assumed. We study within a static approach the stability of a 
   protoneutron star. In particular we focus on a suitable 
   temperature profile suggested by dynamical calculations. We consider a 
   protoneutron star as composed of an isothermal core and an isentropic 
   outer part. 
   To describe physical properties of the interior we employ a microscopically 
   derived equation of state for nuclear matter. For the outer part we 
   employ the Lattimer-Swesty equation of state. The global structure is studied. 
   The assumed temperature profile turns out to give a range of 
   stability which supports temperature values in line with those 
   coming from dynamical calculations. The maximum mass instead depends only upon 
   the equation of state employed.
}



\maketitle

\section{Introduction}
Soon after a supernova explosion a protoneutron star (PNS) is formed and 
constitutes for several tens of seconds a transitional state to 
either a neutron star (NS) or a black hole (BH). In the early stage of 
evolution a PNS is dominated by neutrino diffusion (\cite{borr}) from the core to 
the surface, resulting firstly in deleptonization, which heats the star 
generating temperature values ranging from $30$ to $50$ MeV. \\
In this note we will focus on the study of the structure of a PNS in its 
early stage in the framework of a static approach. We assume, prompted 
by several dynamical calculations (\cite{borr}, \cite{pons}, \cite{suzuki}), 
that in this stage 
the neutrinos are trapped and the temperature is high. In particular we will 
study those configurations where the core has a temperature greater 
than $10$ MeV and the outer part of the star is characterized by a high 
value of entropy per baryon. When one deals with a static description of 
these configurations a problem occurs: what temperature profile shall we 
assume? This key question will be the central point of this note.\\
The basic input for each description of PNS is the nuclear matter 
equation of state (EoS). For the neutrino-rich core the 
non-relativistic Brueckner-Hartree-Fock (BHF) EoS at finite temperature 
(\cite{baldo}) was implemented. We will not give here further details 
on the EoS implemented and we refer the reader to the original works 
(\cite{bf}). For the high-entropy envelope we implement the 
EoS developed by Lattimer-Swesty (LS) (\cite{ls}). A description of PNS within 
the BHF approach has already been given in a preceding work (\cite{aap}). 
In this sense this note can be considered as a refinement of that 
work with focus on a particular choice for the temperature and entropy 
profiles.\\
The note is organized as follows. In section \ref{model} we present 
a new static approach for PNS which is built taking care as much 
as possible of all information coming from dynamical calculations 
mentioned above. Section \ref{result} is devoted to discussing the 
results implied by the approach. Finally in section \ref{conclusion} we 
draw our conclusions.\\
%
%
%
%
\section{Model}\label{model}
Calculations of static models of protoneutron stars should be 
considered as a first step to describe these objects. In principle the 
temperature profile has to be determined via dynamical calculations 
taking into account neutrino transport properly. Many 
static approaches have been developed in the past decade (\cite{bomb}), 
implementing several finite temperature EoS and assuming an 
isentropic or an isothermal (\cite{bomb}) profile throughout the star. 
Nevertheless these simple assumptions on the temperature profile 
lead to some problems. For the case of an isentropic profile static models 
can predict, in pure nucleonic PNS, values of central temperature 
higher than those predicted by dynamical simulations, or they are unable to 
reproduce correctly minimum masses for the isothermal case. For the latter 
case much better results were obtained using a more realistic prescription
(\cite{hean}), which also allows to predict location and temperature of the 
neutrinosphere. A very interesting assumption on temperature and 
entropy profiles can be found also in \cite{strob}, where the entire PNS 
evolution consists of four stages.\\
In our model we assume that a PNS is composed of a hot, neutrino 
opaque, and 
isothermal core separated from an outer cold crust by an isentropic, 
neutrino-free intermediate layer, which will be called the envelope 
throughout the paper. 
%
\subsection{isothermal core}
For a PNS core in which the strongly interacting particles are only baryons, 
its composition is determined by requirements of charge neutrality 
and equilibrium under weak semileptonic processes,
\begin{equation}\label{betapro}
  B_{1}\rightarrow B_{2}+l+\bar{\nu}_{l},\:\: B_{2}+l\rightarrow 
  B_{1}+\nu_{l}\:\: ,
\end{equation}
where $B_{1}$ and $B_{2}$ are baryons and $l$ is a lepton (either an 
electron or a muon). Under the condition that neutrinos are trapped in 
the system, the beta equilibrium equations read explicitly
\begin{equation}\label{betaeq}
  \mu_{n}¥-\mu_{p}=\mu_{e}-\mu_{\nu_{e}}=\mu_{\mu}-\mu_{\nu_{\mu}}\:\: .
\end{equation}
Because of trapping, the numbers of leptons per baryon of each 
flavour ($l=e,\mu$), 
$Y_{l}=x_{l}-x_{\bar{l}}+x_{\nu_{l}}-x_{\bar{\nu}\:\:_{l}}$, are conserved. 
Gravitational collapse calculations of the iron core of massive 
stars indicate that, at the onset of trapping, the electron lepton 
number is $Y_{e}\simeq 0.4$; since no muons are present at this stage we 
can impose also $Y_{\mu}= 0$. For neutrino free matter we just set 
$\mu_{\nu_{l}}=0$ in eq.(\ref{betaeq}) and neglect the above constraints 
on lepton numbers. We assume a constant value of temperature 
throughout the core and perform some calculations for a value of temperature 
ranging from 0 to 50 MeV, with and without neutrinos. The EoS employed 
is that of the BHF approach at finite temperature. Plots for the chemical 
composition and pressure at increasing density and temperature can be 
found in \cite{aap}.
%
   \begin{figure}[t]
   \centering
   \includegraphics[height=9.5cm,clip]{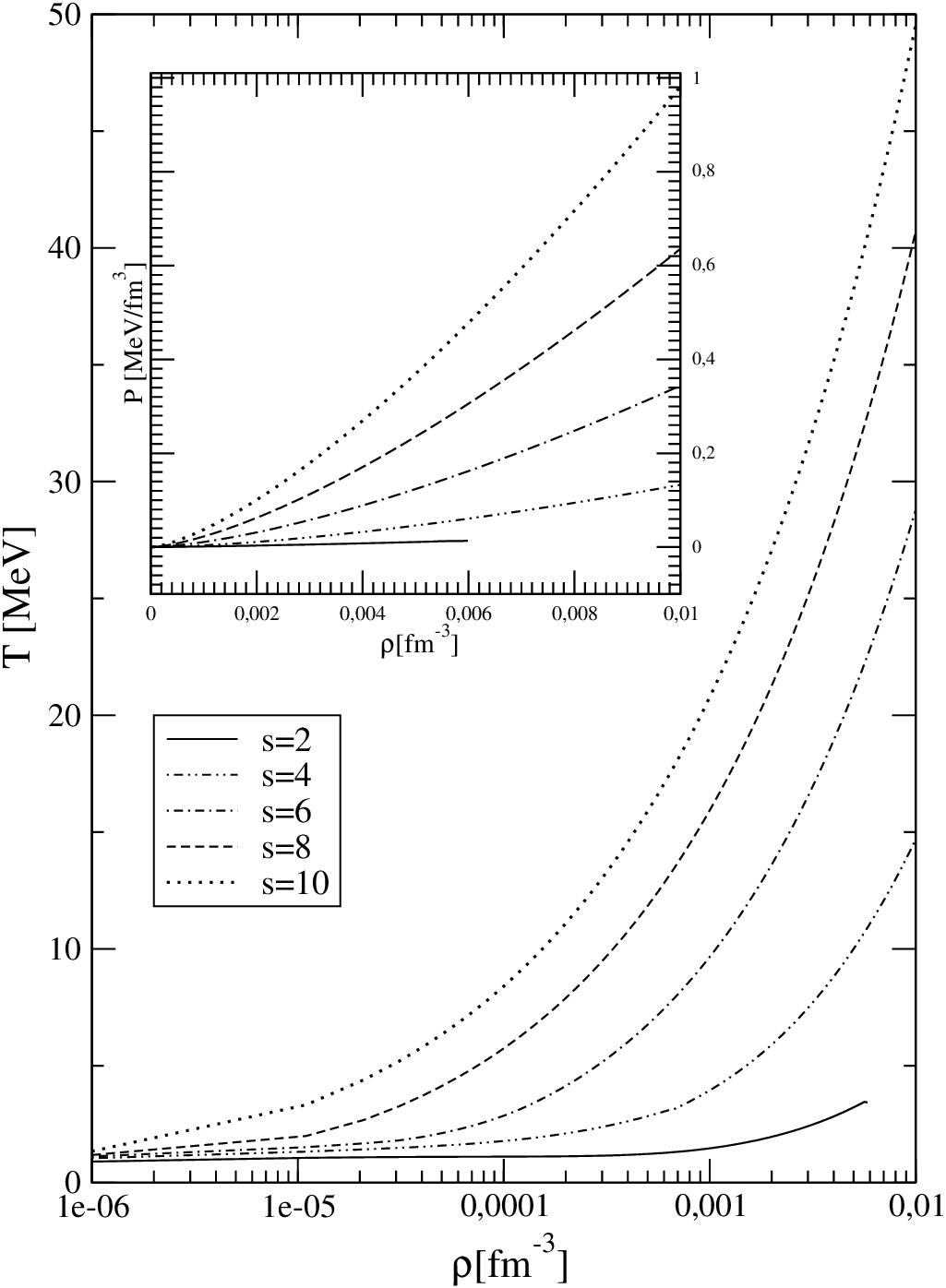} 
      \caption{Temperature profile for different fixed values of entropy per 
   baryon $s$ as a function of the baryon density for the 
   Lattimer-Swesty EoS. The 
   corresponding pressure as a function of baryon density is reported in 
   the inner panel.}
   \label{tprof}
   \end{figure}
%
%
\subsection{isentropic envelope}\label{isen}
The condition of isothermality adopted for the core cannot be 
extended to the outer part of the star. Dynamical calculations 
suggest that the temperature drops rapidly to zero at the surface of the 
star; this is due to the fast cooling of the outer part of the PNS where 
the stellar matter is transparent to neutrinos. Moreover, in the early stage, 
the outer part of a PNS is characterized by a high value of the entropy 
per baryon. Essentially, during the first $10$ seconds, the entropy profile 
decreases from the surface to the core starting from values of 
$6\div 10$ in units of Boltzmann's constant (\cite{borr}, \cite{pons}, 
\cite{suzuki}).\\
For a low enough core temperature ($T\leq10$ MeV) in \cite{aap} a 
temperature profile in the shape of a step function was assumed, 
joining the hadronic EoS (BHF) with the BPS (\cite{bps}) plus FMT 
(\cite{fmt}) EoS for the cold crust. This approach does not work 
anymore if the core temperature $T_{core}$ is greater than $10$ MeV, 
leading to a value for the minimum gravitational mass which rises too 
fast with increasing $T_{core}$. This actually makes the PNS already 
unstable at $T_{core}$ slightly above $30$ MeV in the neutrino-free case.\\
In order to avoid this problem we consider an isentropic envelope in the 
range of baryon density from $10^{-6}$ fm$^{-3}$ to $0.01$ fm$^{-3}$ 
based on the EoS of LS (\cite{ls}) with the incompressibility modulus 
of symmetric nuclear matter $K=220$ MeV (LS220). Fixing the entropy 
per baryon $s$ to 
the above values we get the temperature profiles reported in 
Fig.\ref{tprof}; they explicitly suggest a natural correspondance 
between the entropy of the envelope $s_{env}$ and $T_{core}$. Moreover, 
for the curves displayed in Fig.\ref{tprof} beta equilibrium is 
imposed assuming a neutrino-free regime. 
We thus have the EoS of the envelope with a temperature profile which rises 
quickly to typical values of $T_{core}$, as the density increases, in a 
narrow range of baryon density, just fixing 
the entropy per baryon at a value consistent with dynamical 
calculations.\\ 
Now we need to join together this EoS and that of the star interior.
Since the energy of neutrinos, emerging from the interior, possesses some 
spreading (\cite{bethe}), we do not think that the location of the 
neutrino-sphere, affected by the same spreading, 
is a good criterion to fix the matching density between core and 
envelope. 
Neutrino transport properties indeed vary quite a lot during the PNS 
evolution, they depend on the stellar matter composition 
and change with the leptonic flavour leading to different locations of 
the energy spheres and transport spheres for different neutrino types 
(\cite{janka}).
In addition, the temperature drop regards also those profiles chosen 
as initial conditions for dynamical calculations and borrowed from results on 
iron core collapse simulations, thus establishing its independence 
from the subsequent neutrino diffusion. All this suggests the possibility to 
have a blurring region inside the star where we are free to choose 
the matching density. Conversely, if we consider $s_{env}$ as a free 
parameter in order to join core and envelope energy and free energy densities, 
we recognize soon that only small changes around $\rho_{env}\simeq 
0.01$ fm$^{-3}$ are allowed in order to maintain $s_{env}$ consistent with the 
dynamical expectations. Nevertheless, to build up our model of PNS, a 
fine tuning of $s_{env}$ around the values suggested by 
Fig.\ref{tprof} is performed in order to have an exact 
matching between core and envelope of all the thermodynamic 
quantities (energy density, free energy density, and temperature). In 
this sense we consider $s_{env}$ as a free parameter.
%
   \begin{figure*}[t]
   \centering
    \includegraphics[height=8cm,width=17cm,clip]{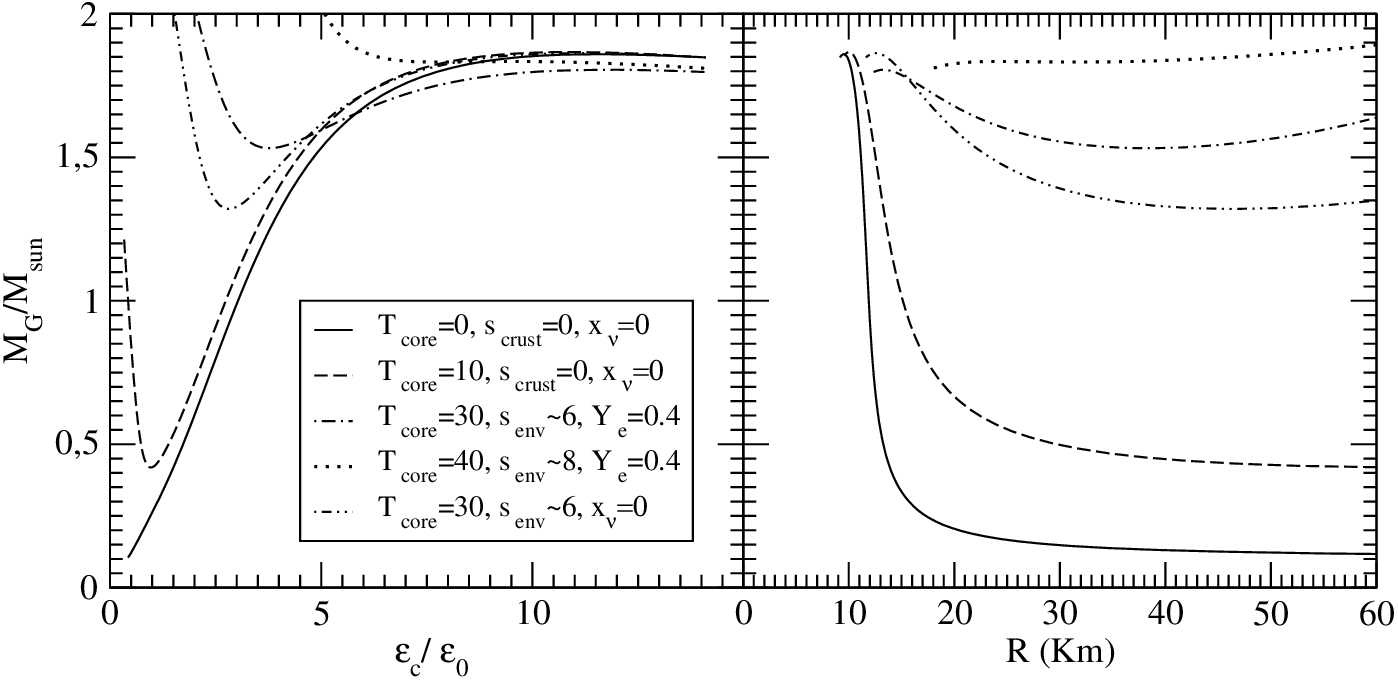} 
   \caption{The gravitational mass $M_{G}$ versus central energy density $\epsilon_{c}$ (left panel) 
   and versus stellar radius $R$ (right panel) for 
   several PNS static configurations (see legend) as discussed in the paper. The 
   gravitational mass is in units of solar masses $M_{\sun}$, central 
   energy density in units of $\epsilon_{0}$ (energy density at 
   saturation), 
   and $R$ in Km.}
   \label{mgr}%
   \end{figure*}
\section{Results}\label{result}
\begin{figure}[t]
   \centering
   \includegraphics[height=8cm,clip]{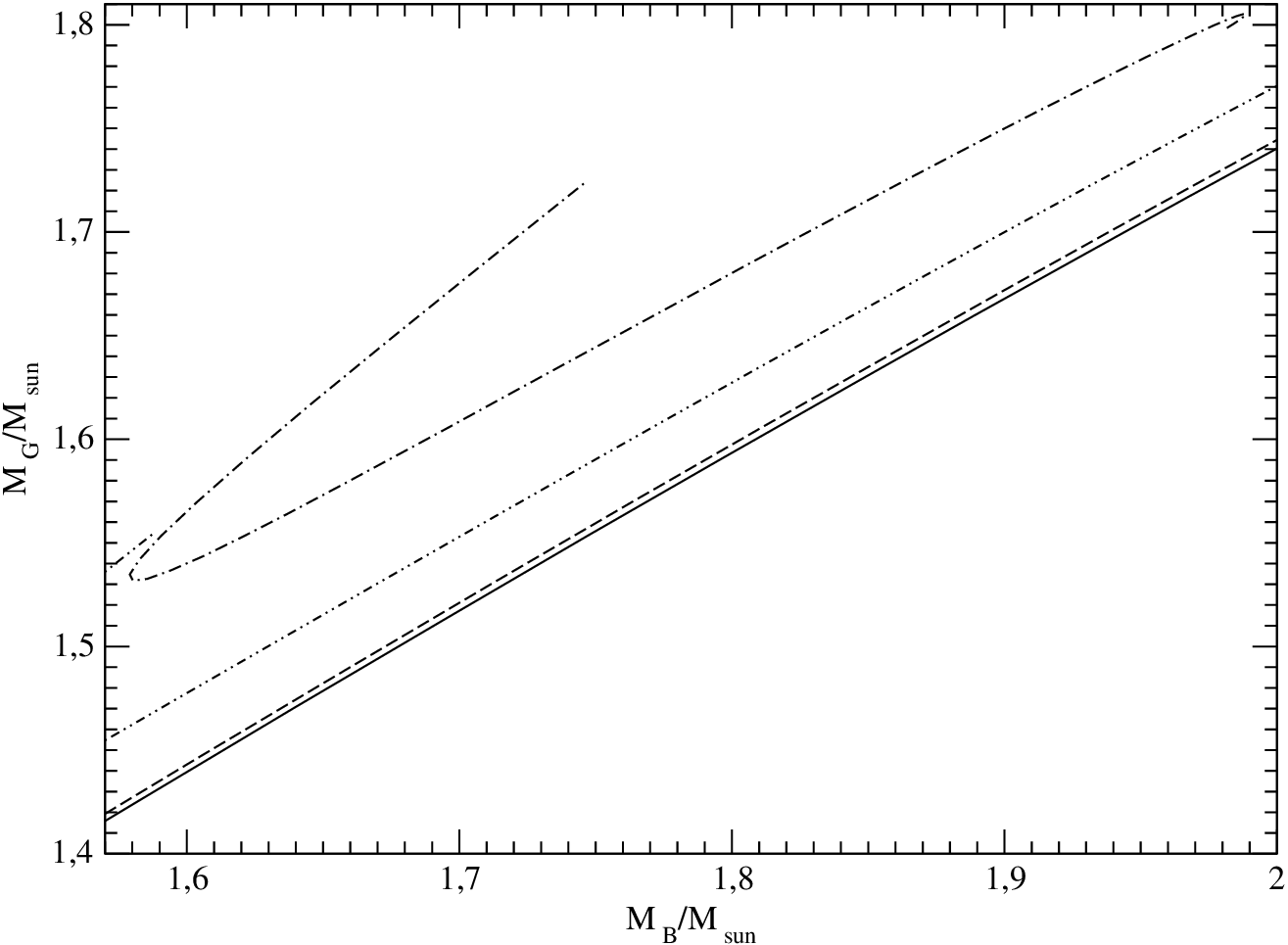} 
      \caption{Gravitational mass $M_{G}$ versus baryonic mass $M_{B}$ 
      for some of the stellar configurations shown in Fig.\ref{mgr}. 
     The notation is the same as in Fig.\ref{mgr} (see legend).}
         \label{mgmb}
   \end{figure}

The stable configurations of a PNS can be obtained from the 
hydrostatic equilibrium equation of Tolman, Oppenheimer, and Volkov 
(\cite{shap}, \cite{zeld}) for the pressure $P$ and the gravitational 
mass $m$,
\begin{equation} 
 \frac{dP}{dr}=-\frac{Gm(r)\epsilon(r)}{r^{2}}
 \frac{\Big[1+\frac{P(r)}{\epsilon(r)}\Big]
 \Big[1+\frac{4\pi r^{3}P(r)}{m(r)}\Big]}{1-\frac{2Gm(r)}{r}}\:\: ,
    \label{tov1}
\end{equation}
\begin{equation}
 \frac{dm}{dr}=4\pi r^{2}\epsilon(r)\:\: ,   
    \label{tov2}
\end{equation}
once the EoS $P(\epsilon)$ is specified, being $\epsilon$ the total 
energy density and G the gravitational constant. For a chosen central value 
of the energy density $\epsilon_{c}$, the numerical integration of Eqs. (\ref{tov1}) 
and (\ref{tov2}) provides the mass-radius and the mass-central 
density relations.\\
We schematize the entire evolution of a PNS as divided in two main 
stages. In the first, representing the early stage, the PNS is in a 
hot ($T_{core}=30\div 40$ MeV) stable configuration with a neutrino-trapped 
core and a high-entropy envelope ($s_{env}\simeq 6\div 8$). 
The second stage represents the end of the short-term cooling where the 
neutrino-free core possesses a low temperature ($T_{core}=10$ MeV) and 
the outer part can be considered as a cold crust (BPS+FMT).\\ 
In Fig.\ref{mgr} we report the gravitational mass as a function of the 
central density (left panel) and stellar radius (right panel) for the 
configurations discussed above. As mentioned before the minimum mass 
plays a fundamental role in determining the range of stability of a PNS 
(\cite{zeld}) and in 
particular it is sensitive of the temperature profile assumed. It is 
found that the introduction of an isentropic envelope yields reliable 
values for the minimum mass for $T_{core}=30$ MeV, as shown in Fig.\ref{mgr}. 
For sake of comparison we also show results for a neutrino-free core at the 
same temperature (dash-double dot line). In the neutrino-trapped case 
the minimum mass rises faster than in the neutrino-free case as the 
temperature increases. This leads to 
unstable PNS at $T_{core}=40$ MeV and $s_{env}\simeq8$ (dotted line). 
This result is consistent with the calculations of \cite{pons} 
where the temperature profiles do not exceed values of $T=40$ MeV.\\
The values for the maximum mass lie within a narrow range $1.77\div 1.87 M_{\sun}$ 
for all the considered configurations. A systematic collection of 
results on maximum and minimum mass together with corresponding radii is 
reported in Tab.\ref{table:1}. The stellar radii at the maximum mass 
(minimum mass) show an overall trend consisting of an increasing 
(decreasing) value as $T_{core}$ increases, both for the neutrino-free and 
trapped case. 
For the second stage the results obtained are very similar to those given 
in \cite{aap} for $T=0,10$ MeV and we do not discuss those results further.\\
In absence of accretion of matter the baryonic mass is constant 
during the evolution of the star and we determine a window of baryonic mass 
that can be supported, which ranges from $1.58 M_{\sun}$ to $1.98 M_{\sun}$ 
for a PNS with $T_{core}=30$ MeV and with trapped neutrinos. In 
Fig.\ref{mgmb} we report 
the gravitational mass $M_{G}$ versus the baryonic mass $M_{B}$ in the 
allowed range. As one can see, for a fixed value of $M_{B}$ the 
gravitational mass decreases as the PNS evolves to a cold NS. This 
difference $\Delta M_{G}$ is on average about $0.1 M_{\sun}$ and 
gives a measure of the energy released during the PNS cooling. For pure 
nucleonic matter metastability\footnote{Here metastability means a 
decreasing stability as long as the PNS evolves to lower values of temperature.} 
does not occur. 
The higher is the temperature of the core the narrower is the window of 
stability for a PNS, therefore in this stage accretion of matter could easily 
lead to a black hole formation.\\ 
The procedure discussed in section \ref{isen} works well even for a PNS with 
hyperons in its interior, giving values for $s_{env}$ in line with 
those for the pure nucleonic case. Since the development of a finite temperature 
BHF EoS with interacting hyperons is still in progress, we prefer to postpone 
this discussion to a future work. 
%
%
%
\begin{table}[t]
\caption{Values of maximum and minimum gravitational mass with their 
corresponding radii and $s_{env}$ for several values of 
$T_{core}$, with and without trapped neutrinos. }             
\label{table:1}      
\centering                          
\begin{tabular}{c c c c c}        
\hline\hline                 
composition & $T_{core}$ & $s_{env}$ & $M^{max}$/$M^{min}$ & 
$R^{max}$/$R^{min}$\\    
     & $[MeV]$ & $[--]$ & $[M_{\sun}]$ & $[Km]$ \\
\hline                        
     & $0$ & $-$ & $1.859/0.106$ & $9.7/174.5$ \\      
 $x_{\nu}=0$ & $10$ & $-$ & $1.867/0.419$ & $10.0/66.8$ \\
     & $30$ & $6.13$ & $1.863/1.320$  & $12.6/46.3$ \\
     & $40$ & $7.87$ & $1.871/1.778$ & $18.7/47.1$ \\
\hline 
     & $0$ & $-$ & $1.769/0.120$ & $9.1/61.4$ \\
 $Y_{e}=0.4$ & $10$ & $-$ & $1.791/0.825$ & $10.3/48.9$ \\
     & $30$ & $6.23$ & $1.805/1.532$ & $13.3/39.4$ \\
     & $40$ & $7.98$ & $-/-$ & $-/-$ \\
\hline                                   
\end{tabular}
\end{table}
%
%
\section{Conclusions}\label{conclusion}
In this paper we have studied the structure and range of stability 
of a PNS within a static description. We have taken into account two main 
stages of the evolution characterized by a hot and neutrino-trapped 
phase and a cold neutrino-free one. The conclusions can be summarized as 
follows.
   \begin{enumerate}
      \item A new temperature profile was assumed which resembles those 
      obtained by dynamical calculations for the first few seconds of 
      life of a PNS. Particular 
      care has been devoted to the description of the outer part where 
      a temperature drop takes place. 
      \item A natural and simple correspondence between temperature of the 
      core and entropy of the envelope has been found. For some aspect 
      a hybrid temperature profile between isothermal and isentropic ones 
      has been adopted allowing us to describe the low central density 
      part of a PNS, which turns out to be important in determining 
      values of the minimum mass.
      \item The minimum mass rises as the temperature of the core 
      increases. This fact becomes particularly evident when the core is 
      neutrino rich, leading to unstable PNS for 
      $T_{core}$ slightly below $40$ MeV.
      \item The maximum mass is much more affected by the composition 
      of the core and therefore by the EoS adopted than by the global thermal 
      effects.
   \end{enumerate}

\begin{acknowledgements}
      The author is very grateful to M.Baldo for many stimulating 
      discussions. He also thanks G.F.Burgio and H.-J.Schulze for 
      their kind help.
\end{acknowledgements}

\end{document}